\def\setup{\count90=0 \count80=0 \count91=0 \count85=0
\countdef\refno=80 \countdef\secno=85 \countdef\equno=90}

\def\autoeql{ {\global\advance\count90 by1} & (\the\count90) }\def\autoeq{ {\global\advance\count90 by1} \eqno(\the\count90) }

\def\autoeql{ {\global\advance\count90 by1} & (\the\count90) }

\def\dddot#1{{\hbox{{\raise 8.5pt\hbox{.}
                     \kern -4pt\raise 8.5pt\hbox{.}
                     \kern -4pt\raise 8.5pt\hbox{.}} \kern -12pt $#1$}}}

\def\autoref{ {\global\advance\count80 by 1} \kern -5pt 
              [\the\count80]\kern 2pt}

\setup
\noindent
\overfullrule = 0pt
\magnification1200
\centerline{  }
\line{\hfill \hbox{cond-mat/9706238}}
\line{\hfill \hbox{Bonn-TH-97-06}}
\line{\hfill \hbox{24th June 1997}}
\vskip 1cm
\centerline{\bf Geometry and Thermodynamic Fluctuations of the}
\centerline{\bf Ising Model on a Bethe Lattice}
\vskip 0.7cm
\centerline{Brian P. Dolan\footnote*{Currently on leave of absence
from: Department of Mathematical Physics, St. Patrick's College, Maynooth,
Ireland}}
\vskip .3cm
\centerline{\it Physikalisches Institut}
\centerline{\it Universit\"at Bonn}
\centerline{\it Germany}
\vskip .3cm
\centerline{e-mail: bdolan@thphys.may.ie}
\bigskip
\vskip 2cm
\centerline{\bf ABSTRACT}
\medskip
\noindent A metric is introduced on the two dimensional space of parameters 
describing the Ising model on a Bethe lattice of co-ordination number $q$.
The geometry associated with this metric is analysed and it is shown
that the Gaussian curvature diverges at the critical point. For the
special case $q=2$ the curvature reduces to an already known result for
the one dimensional Ising model. The Gaussian curvature 
is also calculated for
a general ferro-magnet near its critical point, generalising a previous
result for $t>0$. The general expression near a critical point
is compared with the specific
case of the Bethe lattice and a subtlety, associated with the fact
that the specific heat exponent for the Bethe lattice vanishes, is resolved.
\vskip 1cm
\noindent
\smallskip
\vfill\eject
\leftline {{\bf \S 1 Introduction}}
\bigskip
The concept of defining a distance function, or metric, on the 
space of states of a thermodynamic system or statistical distribution
has recently been investigated by a number of authors (for a review from
the statistical point of view of see
\autoref\newcount\Amari\Amari=\refno, 
from a thermodynamical point of view see 
\autoref\newcount\Ruppeiner\Ruppeiner=\refno). 
The idea has also been extended to field theories 
\autoref\newcount\Zam\Zam=\refno,
\autoref\newcount\HnI\HnI=\refno,
\autoref\newcount\Denjoe\Denjoe=\refno,
\autoref\newcount\geodrg\geodrg=\refno.
For two dimensional field theories a metric 
was used to powerful effect in the proof of the 
c-theorem [\the\Zam].
The purpose of this paper is to investigate further the role of a metric in 
thermodynamics and statistical mechanics by examining 
in detail a class of exactly soluble models --- Ising models on a Bethe 
lattice of arbitrary co-ordination number, $q$. Near the critical point, 
the analysis
should not depend in detail on the underlying model and it is shown how
the specific results for the Bethe lattice fit in with a general analysis
for a ferro-magnet near the critical point.
\bigskip
A metric may be defined on the two dimensional space of states, which can be 
parameterised by the spin-spin coupling $J$ and the external field $H$, or by 
$K=J/T$ and the magnetisation per site $M$.
Near the critical point the curvature can be expressed purely in terms of 
critical exponents and the scaling function, provided the specific heat 
exponent  $0 < \alpha < 1$, and it 
will be shown how the general result for the curvature relates
to that of the Bethe lattice, for which $\alpha =0$.  This generalises the 
result of [\the\Ruppeiner] for the curvature at, and just above, the critical 
temperature with 
vanishing magnetisation.
\bigskip
In the specific example of the Bethe lattice, with co-ordination number q, 
it will be shown that the 
Gaussian curvature is positive and tends to $q/2$ for large temperature, but 
diverges at the critical point $K_c= {{J}\over {T_c}} = {1 \over 2} \ln 
\bigl({{q} \over{q-2}}\bigr)$.  These results generalise those of the one 
dimensional Ising model with periodic boundary conditions derived 
analytically in 
\autoref\newcount\Mrugula\Mrugula=\refno,
and are significant because they are exact results for a class of 
models which exhibits a phase transition.
\bigskip
\leftline {{\bf \S 2 The metric}}
\bigskip
The motivation for a metric comes from probability theory and the theory of 
large deviations 
\autoref\newcount\Ellis\Ellis=\refno.  
For example in thermodynamics [\the\Ruppeiner] 
two states are considered to be close to one another if the probability of 
a fluctuation 
between them is large --- if the probability is unity the distance between the 
states should be zero, if the probability is zero the distance between the 
states should be infinite.  This is satisfied if the distance is a multiple 
of a positive power of minus the logarithm of the probability. The metric
is thus related to the entropy.
\bigskip
Let $g^a$ denote the intensive variables (temperature, pressure, external 
magnetic field, etc.) and $\phi_a$ the corresponding extensive variables.  
The probability of a fluctuation from $\phi_a$ to $\phi_a^\prime$, 
for given values of $g^a$ corresponding to a most probable value $\phi_a$,
is given by a Gaussian distribution for $\delta\phi_a = \phi_a^\prime - 
\phi_a$ small, 
$$
P(\phi^\prime, g) = {\cal N} e^{-{S}(\phi^\prime) + {1 \over 2} 
\bigr\{{{\partial ^2 S (\phi)}\over{\partial \phi_a\partial\phi_b}}\bigl\}
\delta\phi_a \delta\phi_b + \dots} \autoeq$$
where ${\cal N}$ is a normalisation factor and $S$ the thermodynamic 
entropy (the Einstein summation convention is used throughout).
\bigskip
Taking $-\ln P$ we see that a natural metric is given by
$$
dl^2 = - {{\partial^2{S}(\phi)}\over{\partial\phi_a\partial\phi_b}} 
d\phi_ad\phi_b
\autoeq$$
or, in terms of the Legendre transform to $W=-\ln Z$
$$
W(g) = - S(\phi) + g^a\phi_a \autoeq
$$
with $g^a = {{\partial S}\over{\partial\phi_a}}$,
$$
dl^2 = - {{\partial^2 W(g)}\over{\partial g^a\partial g^b}} \delta g^a\delta g^b.
\autoeq$$
\newcount\noncovlineW\noncovlineW=\count90
We shall work on a lattice with $N$ sites and set
$w = {W\over N}={{F}\over{NT}}$, the free energy per site divided by the 
temperature (we use units in which Boltzmann's constant $k=1$).
It will then be convenient to rescale the line element (\the\noncovlineW)
by the constant factor $1/N$ and use
$$
dl^2 = - {{\partial^2 w}\over{\partial g^a\partial g^b}} \delta g^a\delta g^b.
\autoeq$$
\newcount\noncovline\noncovline=\count90
The reason for this rescaling
is in order to ensure that
the metric makes sense in the thermodynamic limit. If the line
element were not rescaled all distances would be infinite in this limit,
since the probability of a fluctuation between two states would always be
zero. The curvature associated with the line element (\the\noncovline) 
is finite in general, except possibly at critical points
(this will be seen explicitly for the Ising model on a Bethe
lattice in the following). Thus the curvature 
associated with the line element
(\the\noncovlineW), which is $1/N$ times that of (\the\noncovline),
would always vanish in the
thermodynamic limit, except at critical points where it would be ill
defined. This procedure of rescaling quantities by powers of $N$
in order to study their behaviour under fluctuations is standard
practice in the theory of large deviations.
Thus the interesting geometry and non-zero curvature associated with 
(\the\noncovline) is due to thermal fluctuations.

Equation (\the\noncovline) is not co-variant under general co-ordinate 
transformations, but that is because it is written in a special co-ordinate 
system.  If $\phi_a = \sqrt{N}<\Phi_a>$ 
is the expectation value of the rescaled extensive quantity 
$\sqrt{N}\Phi_a$, then a co-variant 
form of the line element is  
$$
dl^2 = \{ \langle \Phi_a \Phi_b \rangle - \langle \Phi_a\rangle
\langle\Phi_b \rangle \} dg^adg^b,
\autoeq$$\newcount\covmet\covmet=\count90
which reduces to (\the\noncovline) 
when the partition function is the sum over states of the exponential of a 
{\it linear} combination of $\sqrt{N}\Phi_a$ with co-efficients $g^a$,
$$
Z(g) =  \sum_{\hbox{states}}
e^{-\sqrt{N}g^a\Phi_a}. \autoeq$$\newcount\lineZ\lineZ=\count90
The form of equation (\the\noncovline) is only preserved under 
linear transformations of $g^a$, not under general co-ordinate transformations.
However equation (\the\covmet), which reduces to (\the\noncovline) when the
partition function is of 
the form (\the\lineZ), is general co-ordinate co-variant.  
In what follows it will be convenient to use the special co-ordinate system 
(\the\lineZ) in which the metric takes the form 
$$
G_{ab} = - {{\partial^2 w}\over{\partial g^a\partial g^b}}.
\autoeq$$\newcount\noncovmet\noncovmet=\count90
\bigskip
For systems with two variables, e.g. $K={J\over T}$ and $h={H\over T}$ with 
$J$ a spin-spin coupling and $H$ an external magnetic field,
it is convenient to perform the Legendre transform from $(K,h)$ 
to $(K,M)$ with $M=-{{\partial w}\over{\partial h}}$.  
Changing to these variables, using
$$
G_{a^\prime b^\prime} = {{\partial g^c}\over{\partial g^{a^\prime}}}
{{\partial g^d}\over{\partial g^{b^\prime}}}  G_{cd},
\autoeq
$$
in equation (\the\noncovmet), one finds that the metric in 
the co-ordinate system $(g^{a^\prime}) = (K,M)$ is diagonal,
$$
G_{a^\prime b^\prime} = \pmatrix{ -\Gamma_{KK} & 0 \cr 0&\Gamma_{MM} \cr },
\autoeq$$\newcount\metcomp\metcomp=\count90
where
$$
\Gamma(K,M) = w + hM\qquad\hbox{and}\qquad 
\Gamma_{KK}:={\partial^2\Gamma\over\partial K^2}\qquad
\Gamma_{MM}:={\partial^2\Gamma\over\partial M^2}.
\autoeq$$\newcount\LegTran\LegTran=\count90
Equations (\the\metcomp) and (\the\LegTran) 
provide the basis of the calculations in the following two sections.
\bigskip
\leftline {{\bf \S 3 Curvature of the Bethe lattice}}
\bigskip
The Ising model on a Bethe lattice, of co-ordination number $q$, 
has partition function
$$
Z = \sum_{\{\bar \sigma\}} exp \biggl[K \sum_{(i,j)} \sigma_i \sigma_j + 
h \sum_{i}^{} \sigma_i \biggr ]\autoeq
$$\newcount\ZdefBethe\ZdefBethe=\count90
where $\sum\limits_{\{\bar \sigma\}}^{}$ 
means a sum over all spin configurations on the lattice and 
$\sum\limits_{(i,j)^{}}$ is a sum over nearest 
neighbour pairs, $K = J/T$ is the spin-spin coupling and $h = H/T$ 
an external magnetic field.  A good reference for a description of the
properties of this model is
\autoref\newcount\Baxter\Baxter=\refno, whose notation is used here.
For sites deep inside the lattice the free 
energy per site, $f$, is given by
$$
f/T =- {1\over 2}q K - {1\over 2} q \ln (1 - z^2)
+ {1\over 2} \ln [z^2+1 - z(x+x^{-1})]
+{1\over 2}(q - 2) \ln (x + x^{-1} - 2z) \autoeq$$
\newcount\redfen\redfen=\count90
where $z = e^{-2K}$ and $x$ is defined implicitly by 
$$
x = {{e^{-2K} + e^{-2h} x^{q-1}}\over{1 + e^{-2h-2K} x^{q-1}}}
\autoeq$$\newcount\xdefa\xdefa=\count90
with $0<x<\infty$. This model has a phase transition at $T_c$ given by $K_c = J/T_c = {1\over 2} \ln [q/q-2]$.
The magnetisation per site is
$$
M = {{e^{2h} - x ^q}\over{{e^{2h} + x^q}}}
\autoeq$$\newcount\magx\magx=\count90
which ranges from -1 for $x \rightarrow \infty$ to + 1 for 
$x = 0$. A derivation of equations (\the\redfen)---(\the\magx)
starting from (\the\ZdefBethe) 
can be found in [\the\Baxter].
Explicitly, equations (\the\xdefa) and (\the\magx) give
$$
x(M,z) = {{\sqrt{1-(1-z^2)M^2}} + zM\over{{(1+M)}}} = {{1}\over{x(-M,z)}}.
\autoeq$$\newcount\xdefb\xdefb=\count90
The metric (\the\metcomp) is diagonal if the variables $K$ and $M$ are used, 
so consider the Legendre transform $\Gamma = f/T + Mh$
$$
\eqalign{
\Gamma(K,M) = {1\over 2} q\ln \biggl[ {{2z^2}\over{(1-z^2)}} \biggr] + 
                 {q \over 2} \ln \biggl({{s-z}\over{1-M^2}}\biggr) 
  &+ {1\over 2} \ln\biggl( {{1-M^2}\over{4}}\biggr) + 
     {{qM}\over{2}} \ln \biggl( {{s+zM}\over{1+M}}\biggr) \cr 
&+ {M\over 2} \ln \biggl( {{1+M}\over{1-M}}\biggr)
      }\autoeq$$\newcount\Gamdef\Gamdef=\count90
where $z=e^{-2K}$ and $s=\sqrt{1-(1-z^2)M^2}$.  Equation 
(\the\xdefb) ensures that $\Gamma(K,M) = \Gamma(K,-M)$.  
The derivation of equation (\the\Gamdef) from equation (\the\redfen) 
can be performed by using equations (\the\xdefa) and (\the\magx) to deduce
$$
{{\partial x}\over{\partial M}} \bigg\vert_K 
  = - {{x(x+x^{-1}-2z)^2}\over{2 \{ 2-z(x+x^{-1}) \}}} \autoeq
$$

$$
{{\partial x}\over{\partial K}} \bigg\vert_M 
    =  {{2zx(x-x^{-1})}\over{ \{ 2-z(x+x^{-1}) \}  }}. \autoeq
$$
As described in the introduction, the metric adopted here involves the 
second derivatives of $\Gamma$
$$
G = \pmatrix{ -\Gamma_{KK} & 0 \cr 0&\Gamma_{MM} \cr }.
\autoeq$$\newcount\diagmet\diagmet=\count90
One finds,
$$
\eqalign{
\Gamma_{MM} &= {{(y-2z)^2}\over{4}} \biggl [ {{1}\over{(1+z^2-zy)}} + 
{{q}\over{(zy-2)}} \biggr ]\cr \Gamma_{KK} 
&= 8q \biggl [ {{z^2}\over{(1-z^2)^2}} + {{z}\over{(y-2z)(zy-2)}} \biggr] }
\autoeq$$\newcount\Hessian\Hessian=\count90
where ${y = x+x^{-1} = 2} {{{\{\sqrt{1-(1-z^2)M^2}} - zM^2 \}  
\over{(1-M^2)}}}$.
The Gaussian curvature can be determined from the metric (\the\diagmet)
using (\the\Hessian).
The algebra is tedious, but the final result is remarkably simple
$$
{\cal R} = {(1-z^2)\over 2}{{\{2(1+z^2 -yz) - 
q(3z^2-2zy+1)\}}\over{\{(zy-2) + q(1+z^2-yz)\}^2}}.
\autoeq$$
This can be written explicitly as a function of $M$ and $K$, with $z=e^{-2K}$ 
and $s= \sqrt{1-(1-z^2)M^2}$,
$$
{\cal R}(K,M) =  {{(1-z^2)(1-M^2)}\over{2}}
{{\{(q-2)(3z-s)+4z\}}\over{(s-z)\{(q-2)(z-s)+2z\}^2}}\; .
\autoeq$$\newcount\Gauss\Gauss=\count90
For $T\rightarrow\infty$  $(K=0)$ this reduces to
$$
{\cal R}(0,M) = q/2
\autoeq$$
and for $M^2=1$
$$
{\cal R}(K,1) = q/2.
\autoeq$$
The curvature is plotted in figure 1 for $q=3$, higher $q$ give qualitatively 
similar pictures.
\bigskip
Near the critical point, $T=T_c(1+t)$ with $M<<1$ and $t<<1$,
$z$ is of the form $z=z_c\{1+2K_ct + o(t^2)\}$
with $K_c = {1\over 2} \ln \bigl({{q}\over{q-2}}\bigr)$ and
$z_c = {{q-2}\over{q}}$. One then finds
$$
{\cal R} \approx {1\over 2} {{(q-1)^2}\over{q^2(q-2)}} 
{{1}\over{\bigl\{{1\over 2}\ln\bigl( {{q}\over{q-2}}\bigr)t+ 
\bigl( {{q-1}\over{q^2}}\bigr) M^2\bigr\}^2}}\;.\autoeq$$ 
\newcount\BetheRcrit\BetheRcrit=\count90
In particular, ${\cal R}$ diverges to $+\infty$ 
at the critical point $(M = 0, t \rightarrow 0)$. 
\bigskip
The line of spontaneous 
magnetisation for $t<0$ is given by the smallest solution of
$$
z = x {{(1-x^{q-2})}\over{(1-x^q)}},
\autoeq$$\newcount\spontmag\spontmag=\count90
see reference [\the\Baxter], and ${\cal R}$ is finite along this line,
except at the critical point $M=t=0$.
For small negative $t$, equation (\the\spontmag) 
gives the equation for the spontaneous magnetisation
$$
M^2 = (-t) {{3q^2}\over{2(q-1)}} \ln \biggl( {{q}\over{q-2}}\biggr)+o(t^2).
\autoeq$$
Together with equation (\the\BetheRcrit) this determines the 
Gaussian curvature along the line of spontaneous magnetisation as
$$
{\cal R} = {1\over 2} {{(q-1)^2}\over{q^2(q-2)}}
{{1}\over{\bigl\{ (-t) \ln \bigl({{q}\over{q-2}}\bigr)\bigr\}^2}} 
+ o\biggl({1\over t}\biggr),
\autoeq$$
which is finite except at the critical point $t=0$.
\bigskip
The curvature diverges along the line on which the magnetic susceptibility 
diverges (the pseudo-spinodal curve 
\autoref\newcount\GauntDomb\GauntDomb=\refno). This line lies in the
unstable region and
does not coincide with the line of spontaneous magnetisation 
(except when $q\rightarrow \infty$) --- though they touch at the 
critical point. These two curves are plotted in
figure 2 for the case $q=3$.
\bigskip
These results are in complete agreement with the general analysis in 
[\the\Ruppeiner]. Equation (\the\Gauss) reduces to the known
result for one dimensional Ising model where 
$q=2$ and $M={\sinh h/\sqrt{\sinh^2 h +e^{-4K}}}$,
[\the\Mrugula]. 
\bigskip
\leftline {{\bf \S4 Curvature Near the Critical Point}}
\bigskip
Near the critical point the \lq\lq effective action" for a general 
ferro-magnet
$$
\Gamma(t,M) = {f\over T} + Mh
\autoeq$$
is assumed to have the scaling form, independent of the microscopic
model,
$$
\Gamma=\vert t\vert^{2-\alpha} F(x)\quad +\quad  \hbox{analytic part},
\autoeq$$\newcount\ScalFunc\ScalFunc=\count90
where $x = t/M^{1/\beta}$ is a scaling variable and $F(x)$ a 
scaling function, $\alpha$ and $\beta$ are 
standard critical exponents (see for example
\autoref\newcount\ZinnJustin\ZinnJustin=\refno). 
In the variables $t$ and $M$, 
the metric is diagonal
$$
G = \pmatrix{ 
-{{\partial^2\Gamma}\over{\partial t^2}} & 0 \cr 
0&{{\partial^2\Gamma}\over{\partial M^2}} \cr }.
\autoeq$$\newcount\shmet\shmet=\count90
Thus $(-G_{tt})$ is the specific heat and $G_{MM}$ the inverse magnetic 
susceptibility.
\bigskip
In terms of the scaling function, $F(x)$.
$$
\eqalign{
G_{tt} &=-\vert t \vert^{-\alpha} \{(2-\alpha) (1-\alpha) F 
 + (3-2\alpha) \dot F + \ddot F \}\cr
G_{MM} &= {{\vert t \vert^{2 - \alpha-2\beta}}\over{\beta^2}} x^{2\beta} 
(\beta \dot F + \ddot F)\cr }\autoeq$$\newcount\critmet\critmet=\count90
where $\dot F = x {{dF}\over{dx}}$ etc.  Stability requires $G_{tt} > 0$ 
and $G_{MM} > 0$, which puts conditions on the function $F$,
$$
\eqalign{
(2-\alpha) (1-\alpha) F &+ (3-2\alpha) \dot F + \ddot F < 0 \cr
x^{2\beta}(\beta \dot F &+ \ddot F) > 0. \cr } \autoeq
$$\newcount\stabcond\stabcond=\count90
In order that the internal energy $U = - 
{{\partial \Gamma}\over{\partial t}}$ be finite at $t=0$, 
we must have $\alpha < 1$, thus $(2-\alpha)(1-\alpha)>0$.
For $x>0$ eliminating $\ddot F$ from (\the\stabcond) requires
$$
{{(3-2\alpha-\beta)}\over{(2-\alpha)(1-\alpha)}} \dot F < F.
\autoeq$$
The calculation of the Gaussian curvature from equations 
(\the\critmet) is 
straightforward but tedious.  The final result can be succinctly expressed by 
introducing a \lq\lq dual" 
scaling function ${\cal {F}} (x) = x^{2-\alpha} F(x)$. 
The result is
$$
{\cal R} = {{1}\over{4\vert t \vert^{2-\alpha}}}{{(\alpha + \beta - 1)}\over
{(\beta \dot F + \ddot F)(\ddot {\cal F}-\dot{\cal F}) }}  
\biggl[{{(1-\alpha)(\dddot {\cal F}\dot{\cal F}-\ddot{\cal F}^2)}\over
{(\ddot {\cal F}-\dot{\cal F})}} - {{(2-\alpha-\beta)(\dddot F 
\dot F-\ddot F^2)x^{2-\alpha}}\over{(\beta \dot F + \ddot F)}} \biggr].
\autoeq$$\newcount\Rcrit\Rcrit=\count90
This extends the result of [\the\Ruppeiner] 
to the case of non-zero magnetisation, 
and so is valid for $t<0$ as well as $t>0$, (for 
comparison note that the result of reference [\the\Ruppeiner] 
is quoted in terms 
of a different function, ${{1}\over {T_c}} Y (M/t^{\beta}) = 
F(t/M^{1\over\beta})$).  The derivation of equation (\the\Rcrit) 
assumes that the $\vert t \vert^{-\alpha}$ term in $G_{tt}$ 
(equation (\the\critmet) ) dominates over the analytic part, 
i.e. it assumes that $\alpha > 0$, so it is only valid for 
$0<\alpha<1$.  This means that equation (\the\Rcrit) cannot be 
applied directly to the Bethe lattice at the critical point, for which 
$\alpha=0$.
\bigskip
The scaling function, $F(x)$ in equation (\the\ScalFunc), 
for the Bethe lattice can be obtained by Taylor expanding $\Gamma(K,M)$ 
in equation (\the\Gamdef) about $M=0$ and extracting the co-efficient of 
$t^2$. First expand $\Gamma(K,M)$,
$$
\eqalign{
\Gamma(K,M) &= {q\over 2} \ln \biggl( {{z^2}\over{1+z}}\biggr) + 
{({q-1})\over{2}}  \ln 2 \cr
&+ {{M^2}\over{2!}}\biggl\{ 1 - {{q}\over{2}} (1-z) \biggr \} \cr
&+ {{M^4}\over{4!}} \biggl\{2 - {{q}\over{2}} (z+2)(1-z)^2 \biggr\}\cr
&+ {{M^6}\over{6!}} \biggl\{ 24 - {{3}\over{2}} q (1-z)^3(3z^2+9z+8)\biggr\}\cr
&+ o(M^8). \cr } 
\autoeq$$\newcount\Gexpan\Gexpan=\count90
Now let $t= {{T-T_c}\over{T_c}}$, so that 
$z = e^{-{{2K_c}\over{1+t}}}$ where 
$K_c = {1\over 2}\ln \bigl({{q}\over{q-2}}\bigr)$, giving
$$
\Gamma = D + Et + t^2 F(x) + o(t^3)
\autoeq$$\newcount\lettdefa\lettdefa=\count90
where the scaling function has the form (also derived
in reference [\the\Baxter])
$$ F(x) = a + {{b}\over{x}} + {{c}\over{x^2}},
\autoeq$$\newcount\Fdef\Fdef=\count90
and we have defined the following constants
$$
\eqalign{
D &= {{q}\over{2}} \ln \biggl\{ {{(q-2)^2}\over{q(q-1)}} \biggr\} - \ln 2\cr
E &= {{q}\over{2(q-1)}} (3q-2) K_c\cr}
\autoeq$$\newcount\lettdefa\lettdefa=\count90
$$\eqalign{
a &= -{1\over 4} {{qK_c}\over{(q-1)^2}} \{2(3q-2)(q-1) + q(q-2)K_c\}\cr
b &= {1\over 2}(q-2)K_c\cr
c &= {{(q-2)(q-1)}\over{12q^2}}\cr}\autoeq
$$\newcount\lettdefb\lettdefb=\count90
(remember $x= t/M^{1/\beta}$).
The subtlety here is that, when $\alpha$ is not strictly positive,
(\the\shmet) is not quite correct because it assumes that the
non-analytic term ${1\over t^\alpha}$ dominates over the analytic part,
which is not true when $\alpha=0$.
In general one has
$$
\eqalign{
G_{tt} &= \biggl( {{dK}\over{dt}}\biggr)^2 G_{KK} 
= -{{\partial}\over{\partial t}} \biggl({{dK}\over {dt}} 
{{\partial \Gamma}\over{\partial K}}\biggr) + 
{{d^2K}\over {dt^2}} {{\partial \Gamma}\over{\partial K}}\cr
&= - \Gamma_{tt} + \biggl\{ {{d^2K}\over {dt^2}} 
\biggl({{dK}\over {dt}}\biggr)^{-1} \biggr\} \Gamma_t. \cr } \autoeq$$
For $\alpha>0$ $\Gamma_{tt}$ dominates over the second term on the
R.H.S., but for the Bethe lattice $\alpha=0$, so the second term
must be retained.
Taylor expanding $K = {{K_c}\over{1 + t}}$ gives 
$$
G_{tt} = - \Gamma_{tt} - 2 \Gamma_t + o(t).
\autoeq$$
Using (\the\lettdefa), (\the\Fdef) and (\the\lettdefb)
with $x = t/M^2$ gives ($\beta = {1\over 2}$ for the 
Bethe lattice)
$$
\eqalign{
G_{tt} &= -2 \biggl( a + {{b}\over {x}} + {{c}\over {x^2}} \biggr) 
- {q \over {(q-1)}} (3q-2) K_c + o(t) \cr 
&= -2 \biggl( \tilde a + {{b}\over{x}} + {{c}\over {x^2}} \biggr) 
+ o(t)  \cr } \autoeq$$
where $\tilde a:= - {1 \over 4} {{q^2(q-2)}\over{(q-1)^2}} K_c^2
 =-{b^2\over 12c}$.
\bigskip
The whole effect of having $\alpha=0$ for the Bethe lattice
can thus be absorbed into a redefinition of the constant $a$ in the scaling 
function ((\the\Fdef) and (\the\lettdefb)).  
Equation (\the\Rcrit) can now be applied, with
$$
F(x) = \tilde a + b/x + c/x^2
\autoeq$$
to obtain the scaling form of the Gaussian 
curvature on the Bethe lattice as
$$
\eqalign{
{\cal R} &= - {{b^2}\over{8\tilde a t^2}} {{1}\over{(b+6c/x)^2}} + 
o\biggl( {{1}\over{t}}\biggr) \cr
&= {{(q-1)^2}\over{2q^2(q-2)\bigl\{ K_ct+ {{(q-1)}\over{q^2}}M^2\bigr\}^2} }
+ o\biggl({{1}\over{t}}\biggr)\cr }\autoeq$$
in agreement with equation (\the\BetheRcrit).
\bigskip
\bigskip
\leftline {{\bf \S5 Conclusions}}
\bigskip
In this paper we have explored the geometrical concept of a metric
on the space of states of a thermodynamic system, two states are defined
as being far apart if the probability of a fluctuation between them
is small. In particular a class of models which exhibits a phase
transition, Ising models on a Bethe lattice, has been examined.
\bigskip
Using expectation values to define a metric on the space of states, as in 
equation (\the\covmet), the geometry of the Ising model on a Bethe 
lattice with co-ordination number $q$ 
has been analysed and the Gaussian curvature is given by equation 
(\the\Gauss ), which is the first main result of this paper.
This reduces to the previously known result for the one 
dimensional Ising model [\the\Mrugula] when the co-ordination number $q = 2$.
The curvature is seen to be positive definite for all temperatures
and magnetisations and diverges at the critical point, but is finite at
all other points (in particular at all other points along the curve
of spontaneous magnetisation). The Gaussian curvature tends to the constant
value $q/2$ for $T\rightarrow\infty$ as well as for $M=\pm 1$ (a
particular case of the latter situation is, of course, $T\rightarrow 0$
--- for $T>0$ it can only be obtained in the limit of the external field
$h$ going to infinity).
\bigskip
For a general ferro-magnetic the scaling form of the Gaussian curvature
near the critical point 
has been calculated in terms of the scaling function and it's 
derivatives, (\the\Rcrit ), which generalises the $t>0$, $M=0$ 
result of [\the\Ruppeiner]. This is our second main result. It has been 
shown how this relates to the scaling form of the Gaussian curvature on the 
Bethe lattice, equation (\the\Gauss).
\bigskip
An interesting open problem associated with the above analysis 
is to ascertain whether or not there is a relationship between the
renormalisation flows for the Ising model on a Bethe lattice and the
geometry described here. For the special case of the one dimensional
Ising model it is already known that the renormalisation flow from
$T=\infty$ to $T=0$, along the $M=0$ axis, is a geodesic and that this
is the only renormalisation 
trajectory that is a geodesic [\the\geodrg]. For the more
general case with $q>2$ it is certainly still true that the $M=0$
axis is a geodesic --- this follows from the observation that the
metric (and curvature) are even functions of $M$ and thus invariant
under a change in sign of $M$ --- and this axis
will be a line of renormalisation flow, but it is an open question
as to whether or not any other renormalisation trajectories are
geodesics for the $q>0$ case. This question merits further investigation.
\bigskip
The author wishes to acknowledge support from the Alexander von
Humboldt foundation as well as the hospitality of the
Physikalisches Institut in Bonn where the manuscript was completed.
Some sponsorship was also received from
Baker Consultants Ltd., Ireland, networking specialists
(http://www.baker.ie).
\vfill\eject
\leftline{\bf References}\hfill
\vskip .5cm
\bigskip
\item{[\the\Amari]} S. Amari, {\it Differential Geometric Methods
in Statistics}, Lecture notes in Statistics {\bf 28} (1985) Springer
\smallskip
\item{[\the\Ruppeiner]} G. Ruppeiner, Rev. Mod. Phys. {\bf 67} (1995) 605
\smallskip
\item{[\the\Zam]} A.B Zamolodchikov, Pis'ma Zh. Eksp. Teor. Fiz. 
{\bf 43} (1986) 565
\smallskip
\item{[\the\HnI]}I. Jack and H. Osborn, Nuc. Phys. {\bf B343} (1990) 647
\smallskip
\item{[\the\Denjoe]} D. O'Connor and C.R. Stephens {\it Geometry, The Renormalisation Group
And Gravity}\break
in {\it Directions In General Relativity},\hfill\break
Ed. B.L. Hu, M.P. Ryan Jr., and C.V. Vishveshwava\hfill\break
Proceedings of the 1993 International Symposium,
Maryland, Vol 1, C.U.P. (1993)
\smallskip
\item{[\the\geodrg]} B.P. Dolan, 
Int. J. Mod. Phys. A {\bf 12} (1997) 2413-2424 
\smallskip
\item{[\the\Mrugula]} H. Janyszek and R. Mruga{\l}a, Phys. Rev. {\bf A 39}, 
(1989), 6551
\smallskip
\item{[\the\Ellis]} R.S. Ellis, {\sl Entropy, Large Deviations and 
Statistical Mechanics}\hfill\break
Grundlehren der Mathematischen Wissenschaften {\bf 271} (1985) Springer
\smallskip
\item{[\the\Baxter]} R.J. Baxter, {\sl Exactly Solved Models in Statistical
Mechanics}\hfill\break
Academic Press, N.Y. (1982)
\item{[\the\GauntDomb]} D.S. Gaunt and C. Domb, J. Phys. {\bf C3}
(1970) 1442
\smallskip
\item{[\the\ZinnJustin]} J. Zinn-Justin, 
{\sl Quantum Field Theory and Critical Phenomena} \hfill\break
(2nd edition) (1993) O.U.P.
\smallskip
\vfill\eject
\noindent Fig. 1: Gaussian curvature for the Ising model on a 
Bethe lattice (co-ordination number $q=3$). The curvature diverges at
the critical point $K=(1/2)\ln(3)\approx 0.5493$, $M=0$, but is 
finite everywhere else in the stable region of the $K-M$ plane.
The line of spontaneous magnetisation 
corresponds to the left hand edge of the graph 
(the vertical axis has been truncated at ${\cal R}=100$).
\par\noindent
\includegraphics{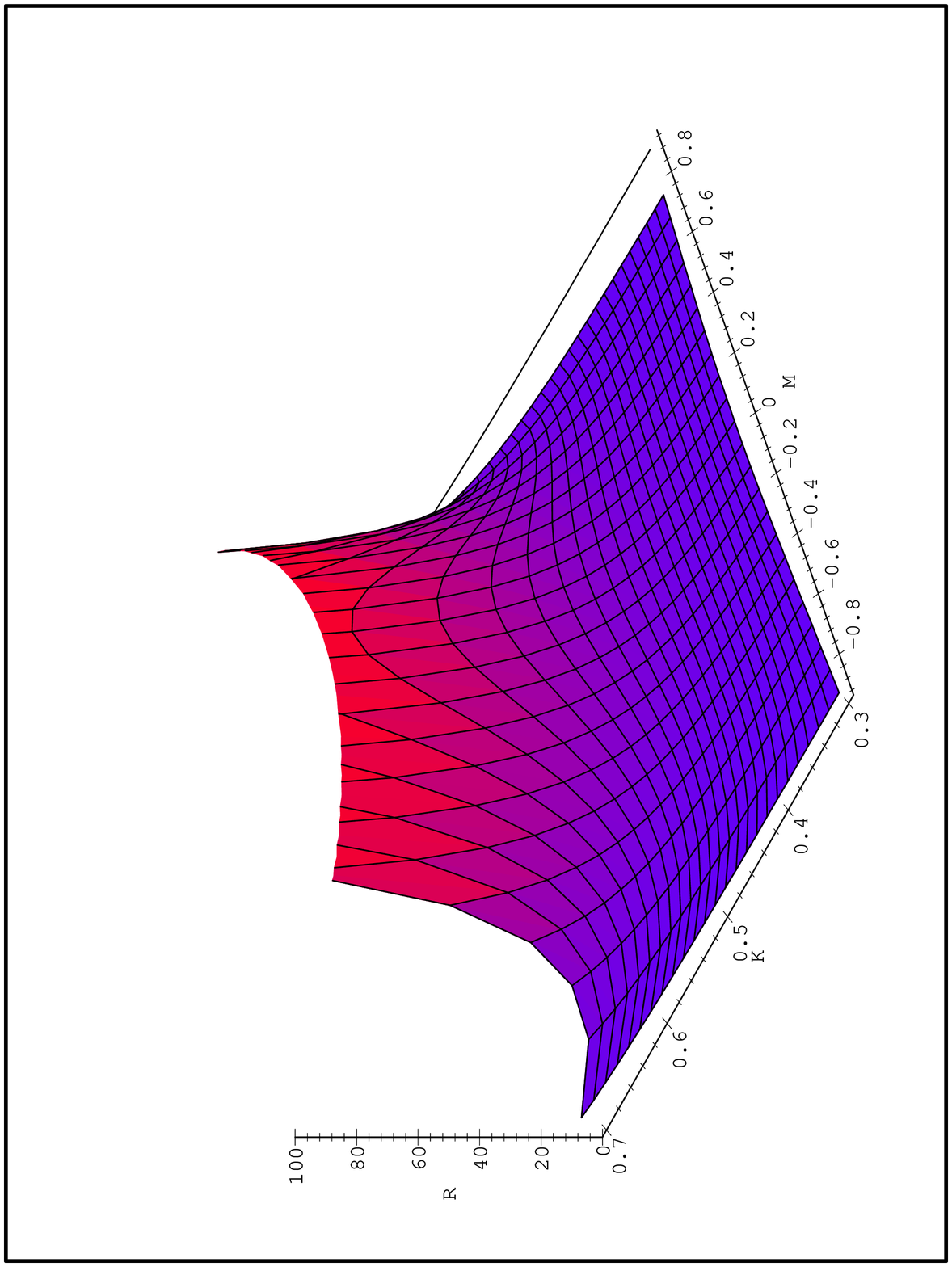}
\vfill\eject
\noindent Fig. 2: Comparison of the line of spontaneous magnetisation
and the pseudo-spinodal
line for the Ising model on a Bethe lattice with co-ordination number $q=3$.
The vertical axis is the magnetisation $M$ and the horizontal axis is 
$z=e^{-2K}$.
The leftmost thick 
line is the pseudo-spinodal curve, along which the susceptibility
diverges (this lies in the unstable region), 
the rightmost thin line is the line of spontaneous magnetisation, 
which corresponds to the left hand  
edge of the 
graph in figure 1. These two lines kiss 
at the critical point $M=0$, $z=1/3$.
\par\noindent
\includegraphics{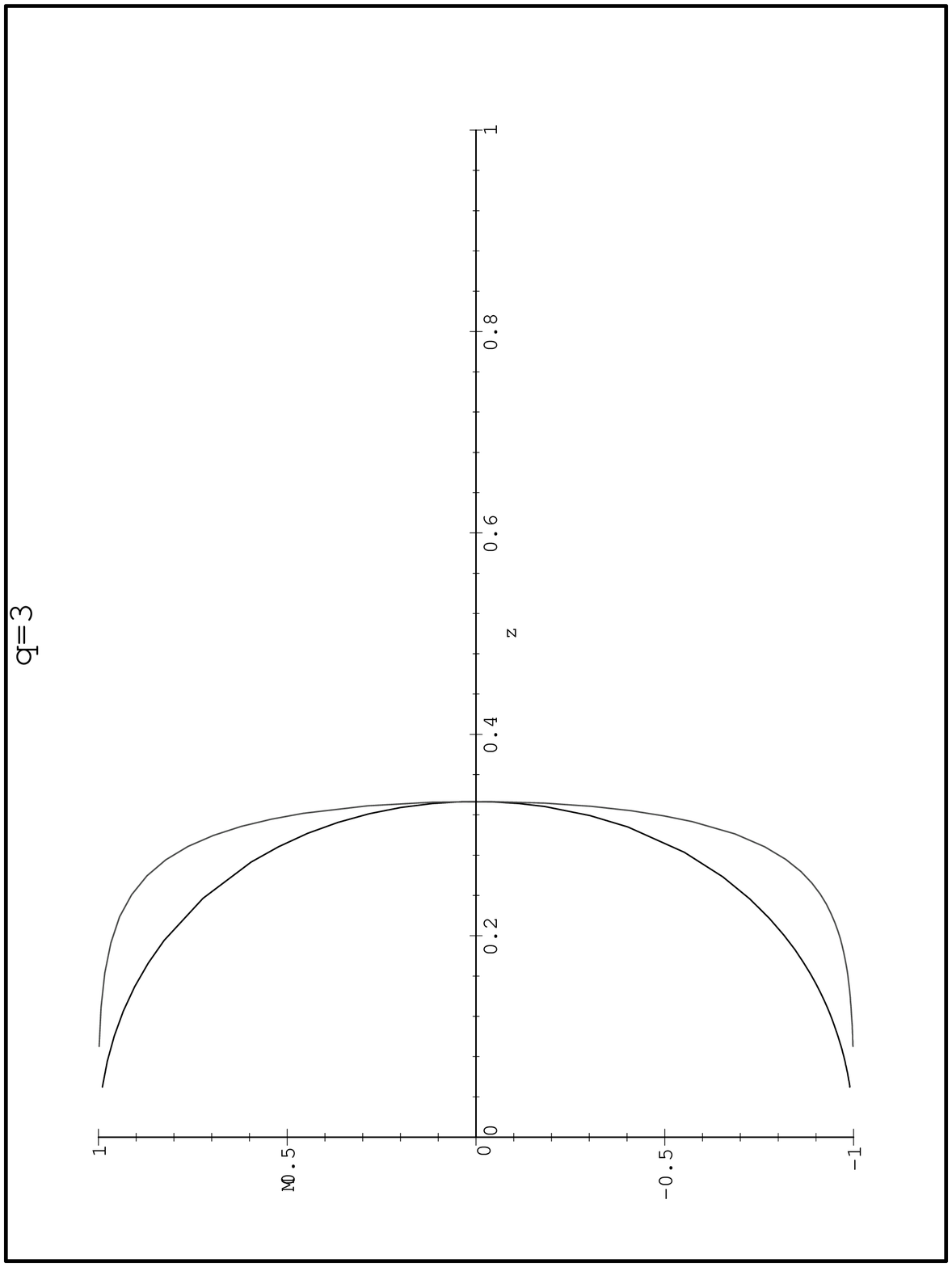}
\vfill\eject
\bye
\end